\documentstyle[12pt]{article}
\input{epsf}
\newlength{\extraspace}
\newlength{\extraspaces}
\newcommand{\be}{\begin{equation}}
\newcommand{\ee}{\end{equation}}
\newcommand{\bea}{\begin{eqnarray}}
\newcommand{\eea}{\end{eqnarray}}

\newlength{\figsize}
\begin{document}
\addtolength{\baselineskip}{.8mm}

\thispagestyle{empty}
\begin{flushright}
{\sc OUPT}-99-16-P\\
hep-th/9903162 \\
March 1999 
\end{flushright}
\vspace{.3cm}

\begin{center}
{\large\sc{ SINGLETONS AND LOGARITHMIC CFT IN ADS/CFT CORRESPONDENCE}}
\\ [12mm]
 {\sc Ian I. Kogan\footnote{i.kogan@physics.ox.ac.uk}}\\
\vspace{.3cm}
{Theoretical Physics, Department of Physics \\[2mm]
 University  of Oxford, 1 Keble Road \\[2mm]
 Oxford, OX1 3NP, UK}
\\[12mm]

{\sc Abstract}
\end{center}
\noindent

We discuss a possible relation between singletons in $AdS$ space and 
logarithmic conformal field theories at the boundary of $AdS$. It is
shown  that the bulk Lagrangian for singleton field (singleton dipole)
induces on the boundary the two-point correlation function for
logarithmic pair. Bulk interpretation of mixing between logarithmic
operator $D$ and zero mode operator $C$ under the scale transformation
is discussed as well as some other issues.            

\newpage
\renewcommand{\footnotesize}{\small}

 The Maldacena conjecture  \cite{M1}
 about duality between  string/M theory on  $AdS_{D+1}$ backgrounds and
 conformal field theory ($CFT_D$) at the boundary of Anti de Sitter  space
 One  particular  example is a duality between $N=4$ supersymmetric 
 Yang-Mills theory
 and  $IIB$ superstrings (supergravity) on $AdS_5\times S^5$ which allows
 to obtain new  exact results in the large $N$ limit of the
strongly coupled superconformal gauge theory. The conformal field
theory   is realized
 as the world-volume theory of $N$ coincident D3-branes  of type $IIB$ string 
theory.  The near horizon geometry of this set of branes is 
$AdS_5\times S^5$ \footnote{for references about near horizon geometry
of p-branes see reviews \cite{horizonreviews}, for a review on
D-branes see \cite{JP}}
 and in the  weak string coupling limit  $g_s \rightarrow 0$
 type $IIB$ theory is described
by a $IIB$ supergravity. The classical description is valid when 
't Hooft coupling  $g^2_{YM}N$ is   large and string loop corrections
become $1/N^2$ corrections.

 This conjecture was further developed in 
\cite{GKP},\cite{W1}  where  very elegant  expression  for  correlation
functions in  $CFT_D$  was given in terms of an action of bulk degrees
of freedom in  $AdS_{D+1}$. A brief review of the $AdS/CFT$
correspondence and of the ideas that led to it formulation is given in
\cite{K1}.
The recipe suggested in \cite{GKP},\cite{W1} for the generating
functional of connected correlation functions in the $CFT_d$ (using as an
example  $N=4$ gauge theory) is to identify it with an extremum (with
the lowest action, when there are several extrema \cite{W2}) of the
classical  supergravity/string action $S[\{\Phi_i\}]$ subject to the boundary
conditions  for bulk fields $\Phi_i$ at the boundary of  $AdS_{D+1}$ 
$\Phi_i(\vec{x}, z = z_{UV}) = \lambda_i(\vec{x})$. 
Here $\vec{x}$ and $z$ are coordinates in $AdS$ and non-zero $z_{UV}$
 plays the role of the
UV cutoff in  a boundary theory\cite{SW}, \cite{GKP}.  These
boundary conditions determine  coupling constants for operators
$O(\vec{x})$  in $CFT_D$ and 
action obtained this way  generates connected correlation functions for
conformal operators $O_i(\vec{x})$  corresponding to bulk fields
 $\Phi_i(\vec{x}, z )$.
The relation between CFT correlation functions and the 
(super)gravity/(super)string bulk action is the following:
\be
 \left< \exp\left[ \sum_{i}\int~ d^Dx ~ 
\lambda_i (\vec{x}) O_i(\vec{x}_i)\right]\right>_{CFT} =  
e^{- S[\{\Phi_i\}]}\arrowvert_{\Phi_i(\delta AdS)  = \lambda_i(\vec{x})}
\label{generatingfunction}
\ee
This relation was  used in many papers to calculate correlation
functions. The 2- and 3-point functions were calculated in \cite{GKP},
\cite{W1}, \cite{three} which gave a spectrum of anomalous dimensions
for different conformal operators. The calculations of 4-point
functions is much more complex and  these functions contain much more
 information about CFT. In several recent  papers 
different aspects of the calculations of 4-point functions in $N=4$ SUSY
 Yang-Mills from $AdS_5$
were discussed \cite{four}, \cite{BG}. In \cite{four} it was done by 
studying exchange diagrams with scalar and gauge fields in the bulk
and in \cite{BG}  using  the $R^4$ term in 
 string corrected type $IIB$ action \cite{GG}. 
It was found that these correlation functions have a surprising property -
there are logarithmic singularities when two operators on the boundary 
approaches each other.  It is necessary to stress here that in both cases 
there were some missing  terms which contributions may cancel logarithmic 
 singularities. In a first case it was a graviton exchange diagram which 
contribution  is still unknown, but hopefully will be found soon
 (see \cite{grav} for a current status) and in the second case there are 
 unknown terms coming from the supersymmetric completion of the $R^4$ term, 
 which also can give non-zero contribution to the four-point functions.

 In this letter we would like to address a question about origin of this 
 logarithmic terms  in case they will not disappear in the final answer.
 Let us consider some $D$-dimensional CFT and assume that in this theory 
 there is a closed operator product expansion (OPE)
 \be
 A(x_1) B(x_2) ~ \stackrel{x_1 \rightarrow x_2}{\longrightarrow}
 ~  \sum_i \frac{f^{i}_{AB}}{|x_1-x_2|^{\Delta_A + \Delta_B - \Delta_i}}
 O_i(x_2)
\label{ope}
\ee 
 Than a  four-point correlation function can be expanded in s-channel 
 $x_{12}, x_{34} \rightarrow 0$, for example, as 
\be
\left<A(x_1) B(x_2) A(x_3) B(x_4)\right> = \sum_{ij} 
\frac{f^{i}_{AB}f^{j}_{AB}}{x_{12}^{\Delta_A + \Delta_B - \Delta_i}
x_{34}^{\Delta_A + \Delta_B - \Delta_j}}\left<O_i(x_2) O_j(x_4)\right>
\label{4point}
\ee
where  two-point functions in the right hand side are non-zero only when
$\Delta_i =\Delta_j$ and 
\be
\left<A(x_1) B(x_2) A(x_3) B(x_4)\right> =  \frac{1}
{x_{12}^{\Delta_A + \Delta_B}
x_{34}^{\Delta_A + \Delta_B}} F(x) 
\label{4point1}
\ee
where $x =  x_{12}x_{34}/x_{24}x_{13}$. In the s-channel $x \rightarrow 0$
 nontrivial part of a correlation function is given by an expansion
 \be
F(x) = \sum_{ij} C_{ij} f^{i}_{AB}f^{j}_{AB} x^{\Delta_i}
 \label{f(x)}
 \ee
where $C_{ij}$ is nonzero only when $\Delta_i =\Delta_j$ and is determined
by two-point functions $<O_i(x) O_j(0)> = C_{ij}/x^{2\Delta_i}$. 
Expansions in  t- and u- channels can be obtained if the OPE is associative
 taking the  limits $x \rightarrow \infty$ and $x \rightarrow 1$. 
  It is easy to see that $F(x)$ has an ordinary power expansion in $x$ and 
can be written as $F(x) = \int d\Delta \rho(\Delta) \exp(-2\Delta lnx)$ where
 density of states 
\be
\rho(\Delta) = \sum_{ij} C_{ij} f^{i}_{AB}f^{j}_{AB}
\delta(\Delta - \Delta_i)
\ee
 To get a   logarithmic singularity at 
 small $x$ one must   have   a rather bizarre  density of 
 states $\rho(\Delta) \sim 1/(\Delta-\Delta_{0})^2, ~ \Delta
\rightarrow \Delta_{0}$,
 i.e. infinitely many states with  dimensions arbitrarily  close to
  some  threshold dimension $\Delta_{0}$.
 It seems that this  is a very exotic behaviour  for 
 any sensible theory  based on AdS/CFT correspondence,
  because   we  have to admit an existence of  infinitely many  fields 
 in the bulk with  practically the same masses. This   leads to
 a lot of disasters, including infinite entropy of the system.  Much
more ``conservative'' approach is to assume that  
 logarithmic terms in $F(x)$ can be explained only if we
 shall assume that among the operators in the r.h.s. of OPE (\ref{ope}) 
 there are {\it logarithmic operators}. These operators
  are an essential feature of the  Logarithmic Conformal Field Theory (LCFT)
 \cite{GUR} and  appear when two (or more) operators $O_i$ in (\ref{ope}) are
 degenerate and have the same anomalous dimensions. In case there are more
  than two degenerate operators, let say $n$,
 there are also terms like $ln^2 x, .., ln^{n-1} x$ in $F(x)$.  For  
 any  pair of degenerate operators $C$ and $D$ (logarithmic pair) the
Hamiltonian becomes  non-diagonalizable \cite{GUR} and acts on logarithmic
 states as
\begin{eqnarray}
L_{0}|C> = \Delta |C>, ~~~~~~ L_{0}|D> = \Delta |D> + |C> \label{example}
\end{eqnarray}
where we used as an example  two-dimensional notations when
 Hamiltonian  is a Virasoro
 operator $L_0$. Degenerate  dimension for logarithmic pair
is  $\Delta$ and 
two-point correlation function \cite{GUR},
\cite{CKT} are given by 
\begin{eqnarray}
\langle C(x) D(y)\rangle &&= 
\langle C(y) D(x) \rangle  = \frac{c}{(x-y)^{2\Delta_C }}\nonumber \\
\langle D(x) D(y)\rangle &&= 
 \frac{1}{(x-y)^{2\Delta_C}} \left(-2c\ln(x-y) + d\right)
\nonumber \\ 
\langle C(x) C(y)\rangle  &&= 0
\label{twopointcorrfunction}
\end{eqnarray}
 The OPE expansion (\ref{ope}) also must be modified
\be
 A(x_1) B(x_2) ~ \stackrel{x_1 \rightarrow x_2}{\longrightarrow}
 ~   \frac{1}{|x_1-x_2|^{\Delta_A + \Delta_B - \Delta}} \left(D + C
\ln|x_1-x_2| + \ldots \right)
\label{logope}
\ee 
The logarithmic terms in (\ref{twopointcorrfunction}) and
(\ref{logope})  immediately lead to $\ln x$ terms in $F(x)$.

 Let us note that our knowledge about LCFT is mostly based on 
two-dimensional  models, where a lot of models  were studied (the
full list of references includes more than 30 papers and full list can
be find in  \cite{lcft}, for example)
 but the way how the two-point functions  were derived in \cite{GUR}, 
\cite{CKT} does not depend actually on $D$ and the structure of  two-point
 correlation functions for logarithmic pair  is  universal for any $LCFT_D$.

  In the AdS/CFT framework the immediate questions to ask  are the 
 following
\begin{itemize}
\item What are the bulk fields corresponding to logarithmic operators ?
      What is so special about these fields (because logarithmic operators
      are definitely very special).
\item How does the Jordan cell structure of logarithmic pair  
   manifest itself in the bulk ?
\item What is the bulk  interpretation of the zero norm of $C$ and mixing
 between $C$ and $D$ under conformal transformations ?
\end{itemize}

 The clue is that a logarithmic pair form an non-decomposable representation
 of conformal algebra (infinite dimensional Virasoro algebra in a 
 two-dimensional case) and  we can look for objects which form
 a   similar non-decomposable representation of $SO(D,2)$ in $AdS_D$. 
 It is amusing that such objects are known for long time - they are
{\it singletons} \cite{singletons},  very special representations 
 corresponding to a "square root" of AdS massless representations. They 
 extension to supersymmetric case, supersingletons were discussed in 
 \cite{supersingletons}.
 In the last year  different aspects of (super)singletons  were discussed
 in relation with AdS/CFT correspondence \cite{adssingletons} but
 without any relation to LCFT.  To see that singletons lead to logarithmic
 correlation functions at the boundary we have to recall that the best
 way to formulate a theory of free  singleton fields in the bulk 
 is in  terms  of a dipole field \cite{dipole} which satisfies
 $ (D^{\mu}D_{\mu} + m^2)^2~A = 0$ or
 \be
 (D^{\mu}D_{\mu} + m^2)~A +  B = 0, ~~~(D^{\mu}D_{\mu} + m^2)~B = 0
\label{dipoleequation}
\ee
and the respective bulk $AdS_{D+1}$ action is
\be
S = \int d^{D+1}x \sqrt{g}\left(g^{\mu\nu}\partial_{\mu}A\partial_{\nu}B  -
 m^2 AB - \frac{1}{2} B^2 \right)
\label{dipoleaction}
\ee
Now we can repeat the same procedure as in \cite{GKP},\cite{W1} and derive two-point
correlation functions for boundary operators $C$ and $D$ corresponding
to dipole pair $B$ and $A$. We do not know yet what pairing it must be
$AC$ and $BD$ or $AD$ and $BC$ and we shall see later  how the choice
 depends on the structure of  Green function matrix. 
  At the boundary of $AdS_{D=1}$ we have
 either coupling $\int d^D x (\alpha A_0 D + B_0 C)$  or 
  $\int d^D x (\alpha A_0 C + B_0 D)$ where $A_0$ and $B_0$ are
boundary values for fields $A$ and $B$. We also introduced a
normalization parameter $\alpha$ which we shall calculate later to
have canonical normalization of logarithmic operators as in 
(\ref{twopointcorrfunction}). The difference between this
case and single scalar fields with mass $m$ is that  instead of one
Green function $K$ which was used as a boundary-bulk propagator we
have a matrix $K_{ij}, ~i,j = 1,2$ now where index $1$ corresponds to
$A$ field and index $2$ corresponds to $B$ field. Using action
(\ref{dipoleaction}  we
can easily get a set of equations for Green functions $K_{ij}$ by a
standard procedure of adding sources and after simple calculations one gets
\bea
(D^{\mu}D_{\mu} + m^2) K_{AA}  + K_{BA} = 0, ~~~~~~~~ 
(D^{\mu}D_{\mu} + m^2) K_{BB} =  0   \nonumber \\
 \label{kij} \\
(D^{\mu}D_{\mu} + m^2) K_{AB} +  K_{BB} =  0, ~~~~~
(D^{\mu}D_{\mu} + m^2) K_{BA}  = 0 \nonumber 
\eea
Now we have  two possibilities. The first one is to assume  
 that Green function must be symmetric $K_{AB} =
K_{BA}$ after which  we can see immediately that $K_{BB}$
 must be zero and
non-dioganal  Green functions solution for this system is a standard 
 boundary-bulk Green function $K$  for a scalar field with mass $m$
which was  discussed in \cite{W1}
\be
K(z,\vec{x};\vec{x}') = \frac{z^{\lambda_{+} + D}}
{(z^2 + |\vec{x}- \vec{x}'|^2)^{\lambda_{+} + D}}
\ee
 where $\lambda_{+} + D = \Delta$ is the dimension of a conformal
field on the boundary. It is determined by a scalar mass $m$ and is
the larger root of a quadratic equation
\be
\Delta(\Delta - D) = m^2, ~~~ \Delta = \frac{1}{2}\left(D + \sqrt{D^2
+ 4m^2}\right)
\label{delta}
\ee
To find $K_{AA}$ we have to solve the equation 
\bea
 (D^{\mu}D_{\mu} + m^2) K_{AA} = - K
\label{kaa}
\eea
and to find a solution  we can use a  following trick. Let us note
that $[d/dm^2,  (D^{\mu}D_{\mu} + m^2)] = 1$, because $AdS$ metric
does not depend on a mass of scalar field propagating in the
bulk. Using the fact that $ (D^{\mu}D_{\mu} + m^2) K =0$ one gets
\be
K = [\frac{d}{dm^2},  (D^{\mu}D_{\mu} + m^2)] K = - (D^{\mu}D_{\mu} +
m^2) \frac{dK}{dm^2}
\ee
 and comparing with (\ref{kaa}) we see that
\be
K_{AA}(z,\vec{x};\vec{x}') =   \frac{dK}{dm^2} = 
-\frac{1}{2\Delta - D} \ln \left(\frac{(z^2 + |\vec{x}-
\vec{x}'|^2)}{z}\right) K(z,\vec{x};\vec{x}') 
\label{main}
\ee
where we used  (\ref{delta}) to get $d\Delta/dm^2 = (2\Delta -
D)^{-1}$. 
Equation (\ref{main}) is  all what we need to get logarithmic
correlation functions on  the boundary. The solution of classical
equation of motions are:
\bea
A(z, \vec{x}) = \int d^D x' \frac{z^{\Delta}}
{(z^2 + |\vec{x}- \vec{x}'|^2)^{\Delta}} B_{0}( \vec{x}') - 
\nonumber \\
\frac{1}{2\Delta - D} \int d^D x' \ln \left(\frac{(z^2 + |\vec{x}-
\vec{x}'|^2)}{z}\right) \frac{z^{\Delta}}
{(z^2 + |\vec{x}- \vec{x}'|^2)^{\Delta}} A_{0}( \vec{x}')  \\
B(z, \vec{x}) = \int d^D x' \frac{z^{\Delta}}
{(z^2 + |\vec{x}- \vec{x}'|^2)^{\Delta}} A_{0}( \vec{x}') \nonumber
\eea

 The second  possibility  arises if we shall not restrict ourselves
 to have a symmetric Green function $K_{ij}$ and the only another
  possible solution of (\ref{kij}) will be 
\bea
K_{BA} = 0, ~~~K_{AA} = K_{BB} = K, ~~~K_{AB} = \frac{dK}{dm^2}
\label{secondchoice}
\eea
 now $K_{ij}$ matrix has a Jordan cell structure with new obvious 
 relations between bulk fields and the boundary values.
 We shall call the first choice  symmetric (S) and the second choice 
 (\ref{secondchoice})-Jordan (J).

 The  classical action can be
evaluated by the same arguments as in the case of a single free scalar
field  using  integration by parts and reduction to a surface
term. Let us show how it can be done in symmetric (S) case.
 Using (\ref{dipoleaction}) and equations of motion
(\ref{dipoleequation}). We shall demonstrate how this can be done for  one can show that action can be expressed as a
surface integral
\be
S(A_0, B_0) = \lim_{\epsilon \rightarrow 0}
\frac{1}{2}\int d^{D}x \sqrt{h}\left(A (\vec{n}\cdot \vec{\nabla})B  +
 B (\vec{n}\cdot \vec{\nabla})A\right)
\ee 
where we take a regularized surface at $z = \epsilon$, normal
derivative is defined as $\vec{n}\cdot \nabla = z
\frac{\partial}{\partial z}$ and  $h$ is an
induced metric on the boundary with $\sqrt{h} = z^{-D}$. One can show
that differentiating the term $\ln(z^2 + |\vec{x}-
\vec{x}'|^2)$ with respect to $z$ will give us sub-leading term
$O(\epsilon)$  and  after straightforward calculations one gets
\bea
S[A_0, B_0] = ~ \Delta \int  d^{D}x \int d^{D}x' \frac{A_{0}( \vec{x})
B_{0}( \vec{x}')}{ |\vec{x}- \vec{x}'|^{2\Delta}}~ - \nonumber \\
\\
 \frac{\Delta}{2\Delta - D} \int  d^{D}x \int d^{D}x' ~
\left[ \ln \left(\frac{|\vec{x}-\vec{x}'|^2}{\epsilon}\right) -
\frac{1}{\Delta}\right]~
\frac{A_{0}( \vec{x})
A_{0}( \vec{x}')}{ |\vec{x}- \vec{x}'|^{2\Delta}} 
 \nonumber
\label{final}
\eea
This is the final result. Using (\ref{generatingfunction}) we can now
immediately extract the two-point functions for conformal fields $C$
and $D$ and now it clear why logarithmic operator $D$ must be connected with
 field $A$ and zero mode operator $C$ must be paired with a
Nakanishi-Lautrup field $B$. We have
\bea
\langle C(\vec{x}) C(\vec{y})\rangle  =  
\frac{\delta^2 S[A_0, B_0]}{\delta B_0(\vec{x}) ~\delta
B_0(\vec{y})} = 0 \nonumber \\
\alpha \langle C(\vec{x}) D(\vec{y})\rangle = 
  \frac{\delta^2 S[A_0, B_0]}{\delta B_0(\vec{x}) ~\delta
A_0(\vec{y})}  = \frac{\Delta}{ |\vec{x}- \vec{y}|^{2\Delta}}\nonumber
\\
\\
\alpha^2\langle D((\vec{x}) D(\vec{y})\rangle  = 
\frac{\delta^2 S[A_0, B_0]}{\delta A_0(\vec{x}) ~\delta
A_0(\vec{y})} = \nonumber \\
 \frac{\Delta}{2\Delta - D}\frac{1}{ |\vec{x}- \vec{y}|^{2\Delta} } 
\left[- 2 \ln \left(\frac{|\vec{x}-\vec{x}'|^2}{\epsilon}\right) +
\frac{1}{\Delta}\right]  
 \nonumber
\label{final2point}
\eea
The standard normalization (\ref{twopointcorrfunction}) can be
obtained after choosing the   factor
$\alpha  = 1/(2\Delta - D)$ which  gives us immediately the constants
$c$ and $d$ in (\ref{twopointcorrfunction})
\be
c = \frac{\Delta}{\alpha} = \Delta(2\Delta -D) = \Delta^2 + m^2, ~~~~
d = \frac{c}{\Delta} = 2\Delta - D = \sqrt{D^2 + 4m^2}
\ee
 Let us note that limiting case
$\Delta = D/2$  arises when $m^2 = - D^2/4$ and this is the lower
bound for the stable scalar filed in $AdS_{D+1}$. We see that in this
case all logarithmic correlation functions become null, because 
$c = d = 0$ and at the same time coupling of logarithmic operator $D$
to field $A$ becomes infinite.

In this letter we shall not present  similar calculations for $J$
case but it is easy to see that we shall get the same structure of the
two-point correlation functions, only with an opposite pairing
 $\int d^D x ( A_0 C + \alpha B_0 D)$ so now it is field $A$ which is
coupled  to a zero-mode operator $C$. The same parameter $\alpha$
stands in front of $D$ operator in both cases and in critical case
$\Delta = D/2$ it is infinite coupling of logarithmic operator $D$ to
field $B$.

 Here we considered only bosonic case, one can study supersingletons
and get the  logarithmic superconformal theory, the details  will be
presented in a future publication.

Let us note that a very similar action (the only difference is the
fixed coefficient in front of $B^2$ term), but without any relation to
singletons  was recently studied in paper \cite{GKA} in which logarithmic
 operators on the boundary were also found. However in that paper only
$J$  case was considered with non-symmetric Green function matrix
$K_{ij}$.   In a subsequent paper \cite{KaG}
fermions were included in the construction using  non-commutative
geometry. It will be interesting to see if there is a connection
between  non-commutative geometry and  supersingletons.

 After we have established a link between singletons in $AdS_{D+1}$
 and $LCFT_{D}$ at the boundary of  $AdS_{D+1}$  and  found an answer
on the first question we can try to answer the questions about bulk 
interpretation of  mixing between $C$ and $D$, zero norm of $C$ and
Jordan cell structure.
 The conformal field theory was defined on a
 regularized surface at $z = \epsilon$.
 Making   a scale transformation
\be
z  \rightarrow z' = z e^{t}, ~~~ \epsilon \rightarrow \epsilon e^{t} 
\label{fsscaling}
\ee 
 it is easy to see  from the scale dependence of the correlation functions
(\ref{final2point}) that $ C_{\epsilon}$ and $D_{\epsilon}$ transform as:
 \bea
D_{\epsilon} &\rightarrow& D_{\epsilon'} =
 D_{\epsilon} - t C_{\epsilon} \nonumber  \\
 C_{\epsilon} &\rightarrow& C_{\epsilon'}= C_{\epsilon}
 \eea
We have to treat $S$ and $J$ cases separately. In the $S$ case the 
coupling is  $\int d^D x (\alpha A_0 D + B_0 C)$  such transformation of
operators $C$ and $D$  corresponds to the following transformation of
 singleton  dipole
 \bea
B_{\epsilon} &\rightarrow& B_{\epsilon'} =
 B_{\epsilon} + \alpha t A_{\epsilon} \nonumber  \\
 A_{\epsilon} &\rightarrow& A_{\epsilon'}= A_{\epsilon}
 \eea
It is unclear if such a transformation of the field corresponds to any
 particular symmetry of the bulk action and it is an interesting open
question. In the $J$-case the coupling is of the form
 $\int d^D x ( A_0 C + \alpha B_0 D)$ and one can get another
transformation law for $A$ and $B$ fields 
\bea
A_{\epsilon} &\rightarrow& A_{\epsilon'} =
 A_{\epsilon} + \alpha t B_{\epsilon} \nonumber  \\
 B_{\epsilon} &\rightarrow& B_{\epsilon'}= B_{\epsilon}
 \eea
But this is precisely the symmetry of action (\ref{dipoleaction})
which variation under the transformation $A \rightarrow A + c B, ~
 B \rightarrow B$ is proportional to equation of motion and zero
on-shell. One can easily see this as a symmetry of equations of 
motion (\ref{dipoleequation}) - operator $D^{\mu}D_{\mu} + m^2$ will 
annihilate shift in $A$ proportional to $B$ due to the equation of
motion for $B$ field. Actually this symmetry is nothing but part of
BRST symmetry of the singleton action after gauge fixing \cite{dipole}.

Let us now address an  issue what does singleton  describe. It was suggested
 \cite{deformed} that they describe the small fluctuations of a
 brane at  the  boundary of the AdS space, i.e. in case of AdS/CFT
 correspondence it is a D-brane. If it is so there may be a very
unusual interplay between two different LCFT. The first one is the
 conformal field theory on a boundary we have just discussed. The
second one may be a world-sheet conformal field theory describing
(super)string in a $D$-brane background. It was suggested some time
ago \cite{KM} that  the world-sheet 
description of the collective coordinates of a soliton in string theory
 is given by logarithmic operators. The problem of $D$-brane recoil
was studied from this point of view in \cite{recoil}.  It  is an extremely  
interesting question if there is any connection between these two LCFT
(let us note that the first LCFT is not two-dimensional in general).
 To do this  one must know a  full  superstring theory in AdS space
with RR background.  In a recent paper \cite{bvw} $AdS_3 \times S^3$
background was studied and a world-sheet conformal field theory was
obtained which was based on a supergroup $SU'(2|2)$. The bosonic
sector of this theory has a Kac-Moody algebra $SU(2)_k \times
SU(2)_{-k}$ and in principle theories with negative $k$  and SUSY WZNW
models often have logarithmic operators in they spectrum ( see
\cite{KM}, \cite{lewis} and references therein).  It will be
interesting to see if these logarithmic operators will survive in the
full superstring spectrum and if they really describe the deformation
of the boundary as was conjectured in \cite{KM}. 
 
 Another interesting question is do we have any place for logarithmic 
operators in $N=4$ Yang-Mills ? In a recent paper \cite{ps} a very
strange behaviour of  $N=4$ Yang-Mills theory in the limit of small
AdS curvature was found  from the  hollographic connection between
gauge fields and gravity. The resolution of paradoxical behaviour
suggested in this paper was based on an assumption that there are a
lot of hidden degrees of freedom which can store information but can
not carry energy density. It seems that the zero norm field $C$ is
well suited for this function and one can think about possible
relation between these conjectured degrees of freedom and logarithmic 
operators. Another interesting issue is possible relation between
logarithmic operators and vacuum instability in the bulk, which was
found in another example of holography -  gauge theories with
Chern-Simons term in three dimensions and two-dimensional LCFT
\cite{lewis2}. In this short letter we do not have place to discuss
these problems. These and other interesting questions about
singletons, topological field theories, vacuum instabilities,
holography and all that will be discussed in a future.

I would like to thank H. Liu, A. Tseytlin and Sanjay for interesting 
discussions and valuable comments about logarithmic terms in 4-point
 correlation functions \cite{four}.


\begin{thebibliography}{99}
\bibitem{M1} J.M. Maldacena, Adv. Theor. Math. Phys. {\bf 2} (1998)
231,  hep-th/9711200.

\bibitem{horizonreviews}
M.J. Duff, R.Khuri, J.X. Lu, Phys. Rep. {\bf 259} (1995) 213;\\
K.S. Stelle,  {\it ``BPS Branes in Supergravity: 
ICTP Summer School  Lectures''}, hep-th/9803116.

\bibitem{JP}
J. Polchinski, Phys. Rev. Lett. {\bf 75} (1995) 4724, hep-th/9510017;
 {\it ``TASI lectures on D-branes''}  hep-th/9611050.
 
\bibitem{GKP} S.S. Gubser, I.R. Klebanov and A.M. Polyakov,
 Phys. Lett. {\bf B428} (1998) 105,  hep-th/9802109.

\bibitem{W1} E. Witten, Adv. Theor. Math. Phys. {\bf 2} (1998) 253,
 hep-th/9802150.

\bibitem{K1} I.R. Klebanov, Talk at Orbis Scieniae '98, December 1998,
 hep-th/9901018.

\bibitem{W2} E. Witten, Adv. Theor. Math. Phys. {\bf 2} (1998) 505,
 hep-th/9803131.

\bibitem{SW} L. Susskind and  E. Witten, hep-th/9805114.

\bibitem{three}
I. Ya. Aref'eva and I.V. Volovich, hep-th/9803028; \\
M. Henningson and K. Sfetsos, Phys.Lett. {\bf B431} (1998) 63,
   hep-th/9803251; \\
W.~M\"{u}ck and K.S.~Viswanathan, Phys. Rev.{\bf D58} (1998) 041901,
 hep-th/9804035; Phys. Rev.{\bf D58} (1998) 106006,  hep-th/9805145; 
  hep-th/9810151\\
D.Z.~Freedman,~S.D.~Mathur,~A.~Matusis and L.~Rastelli,~hep-th/9804058; \\
H.~Liu and A.A.~Tseytlin, Nucl. Phys. {\bf B 533} (1998) 88,hep-th/9804083;\\
T. Banks and M.B. Green,JHEP 9805 (1998) 002,hep-th/9804170; \\
G. Chalmers, H. Nastase, K. Schalm, R. Siebelnik, hep-th/9805105;\\
V. Balasubramanian, P. Kraus, A. Lawrence, hep-th/98051171; \\ 
S.~Lee,~S.~Minwalla,~M.~Rangamani,~and~N.~Seiberg,
~Adv.Theor.Math.Phys. {\bf 2} (1998) 697, hep-th/9806074;\\
E.~D'Hoker,~D.Z.~Freedman,~and~W.~Skiba,~
 Phys.Rev. {\bf D59} (1999) 045008, hep-th/9807098;\\
G. Arutyunov and S. Frolov, hep-th/9901121; \\
F.~Gonzalez-Rey, B.~Kulik and I.Y.~Park, ~ hep-th/9903094.

\bibitem{four}
D.Z.~Freedman, Strings'98 lecture, 
   http:/www.itp.ucsb.edu/online/strings98/;\\
H.~Liu and A.A.~Tseytlin, Phys.Rev. {\bf D59} (1999) 086002,
   hep-th/9807097; \\
D.Z.~Freedman,~S.D.~Mathur,~A.~Matusis and L.~Rastelli,hep-th/9808006; \\
E.~D'Hoker,~D.Z.~Freedman, hep-th/9809179; hep-th/9811257; \\
G.~Chalmers and K.~Schalm, hep-th/9810051; \\
H.~Liu ,hep-th/9811152 . 

\bibitem{BG}
J. H. Brodie and M. Gutperle, Phys.Lett. {\bf B445} (1999) 296,
 hep-th/9809067.

\bibitem{GG}
 M. B. Green and M. Gutperle, Nucl. Phys. {\bf B 498} (1997) 195, 
 hep-th/9701093.

\bibitem{grav}
E. D'Hoker, D. Z. Freedman, S. D. Mathur, A. Matusis, L. Rastelli,
 hep-th/9902042.

\bibitem{GUR} V. Gurarie, Nucl. Phys. {\bf B410} (1993), 535.

\bibitem{CKT} J. S. Caux, I. I. Kogan and A. Tsvelik,
Nucl. Phys. {\bf B466} (1996), 444; hep-th/9511130.

\bibitem{lcft} 
 M. A. Flohr, Phys.Lett. {\bf  B444} (1998), 179;   hep-th/9808169.

\bibitem{singletons} 
 P.A.M. Dirac, J. Math. Phys. {\bf 4} (1963) 901; \\
 M. Flato and C. Fronsdal, Lett. Math. Phys. {\bf 2} (1978) 421,
  Phys. Lett. {\bf B 97} (1980) 263, J. Math. Phys. {\bf 22} (1981) 1100;\\
 for a review of the developments  in the past 20 years:
 M. Flato, C. Fronsdal, D. Sternheimer, hep-th/9901043.

 \bibitem{supersingletons}
 C. Fronsdal,Phys.Rev. {\bf D26} (1982) 1988;\\
 I. Bars and M. Gunaydin, Commun. Math. Phys. {\bf 87} (1982) 159;
{\bf 91} (1983) 21;\\
 M.P. Blencowe and M.J. Duff, Phys. Lett. {\bf B203} (1988) 229;\\
 H. Nicolai, E. Sezgin and Y. Tanii, Nucl. Phys. {\bf B305} (1988) 483;\\
 E. Bergshoeff, A. Salam, E. Sezgin and Y. Tanii, 
  Nucl. Phys. {\bf B305} (1988) 497; Phys. Lett. {\bf B205} (1988) 237;\\
 M. Gunaydin and D. Minic, hep-th/9802047; M. Gunaydin, hep-th/9803138. 

 \bibitem{adssingletons} 
 S. Ferrara and C. Fronsdal, hep-th/9712239, hep-th/9802126, 
 hep-th/9806072;\\
 S. Ferrara, C. Fronsdal and A. Zaffaroni, hep-th/9802203; \\
 S. Ferrara and A. Zaffaroni, hep-th/9803060,hep-th/9807090

 \bibitem{dipole} M. Flato and C. Fronsdal, Commun. Math. Phys. {\bf 108}
 (1987) 469; Phys. Lett. {\bf B 189} (1987) 145
 
\bibitem{GKA} A.M. Ghezelbash, M. Khorrami and A. Aghamohammadi,
hep-th/9807034.

\bibitem{KaG} K. Kaviani and  A.M. Ghezelbash, hep-th/9902104.
 
\bibitem{deformed}
M.J. Duff, {\it Class. Quant. Grav.} {\bf 5}, 
(1988) 189; \\
M. Blencowe and M.J. Duff, {\it Nucl. Phys.} {\bf B310}, (1988) 587.
M.P. Blencowe and M.J. Duff, {\it Phys. 
Lett.}
{\bf B203}, (1988) 229;
E. Bergshoeff, E. Sezgin and Y. Tanii, trieste 
preprint, IC/88/5, (1988);
E. Bergshoeff, A. Salam, E. Sezgin and Y. Tanii, {\it Nucl. Phys.}
{\bf B305}, (1988) 497; H. Nicolai, E. Sezgin and Y. Tanii,{\it Nucl. Phys.}
{\bf B305 \ [FS23]}, (1988) 483.\\
 G. Dall'Agata, D. Fabbri, C. Fraser, P. Fre, P. Termonia,
M. Trigiante, hep-th/9807115, hep-th/9903041

\bibitem{KM} I.I. Kogan and N.E. Mavromatos,  Phys. Lett. {\bf B375}
 (1996), 111;  hep-th/9512210.

\bibitem{recoil}  
 V. Periwal and O. Tafjord, 
Phys.Rev. {\bf D54} (1996) 3690, hep-th/9603156;\\
 D. Berenstein, R. Corrado, W. Fischler, S. Paban
and M. Rozali, Phys.Lett. {\bf B384} (1996) 93,  hep-th/9605168;\\
I. I. Kogan, N. E. Mavromatos and J. F. Wheater,
Phys.Lett. {\bf B387} (1996) 483, hep-th/9606102;\\
 J. Ellis, N.E. Mavromatos and D.V. Nanopoulos, 
Int.J.Mod.Phys. {\bf A13} (1998) 1059, hep-th/9609238.

\bibitem{bvw} N. Berkovits, C. Vafa and  E. Witten, hep-th/9902098.

\bibitem{lewis}  J.-S. Caux, I. Kogan, A. Lewis, A. M. Tsvelik,
 Nucl.Phys. {\bf B489} (1997) 469, hep-th/9606138; \\
I. I. Kogan and  A. Lewis,  Nucl.Phys. {\bf B509}
 (1998) 687, hep-th/9705240; \\
I. I. Kogan,  A. Lewis and O. A. Soloviev, 
Int.J.Mod.Phys. {\bf A13}  (1998) 1345,  hep-th/9703028; \\
 J.-S. Caux, N. Taniguchi and  A. M. Tsvelik, Nucl.Phys. {\bf B525}
 (1998) 671,    cond-mat/9801055.

\bibitem{ps} J. Polchinski and L.  Susskind, hep-th/9902182

\bibitem{lewis2} I. I. Kogan and  A. Lewis, Phys.Lett. {\bf B431} (1998)
77, hep-th/9802102; \\
A. Lewis, Nucl.Phys. {\bf B539} (1999) 367,  hep-th/9808068.


\end{thebibliography}
\end{document}